\documentclass[aps, prl, reprint, superscriptaddress, floatfix]{revtex4-1}
\usepackage{amsmath, amssymb, graphicx, bbm}
\usepackage[hidelinks]{hyperref}
\usepackage{subfigure}

\begin{document}

\title{Theory of Two-Dimensional Nonlinear Spectroscopy for the Kitaev Spin Liquid}

\author{Wonjune Choi}
\affiliation{Department of Physics, University of Toronto, Toronto, Ontario M5S 1A7, Canada}
\author{Ki Hoon Lee}
\affiliation{Center for Correlated Electron Systems, Institute for Basic Science (IBS),
Seoul National University, Seoul 08826, Korea}
\affiliation{Department of Physics and Astronomy, Seoul National University, Seoul 08826, Korea}
\author{Yong Baek Kim}
\affiliation{Department of Physics, University of Toronto, Toronto, Ontario M5S 1A7, Canada}

\begin{abstract}
Unambiguous identification of fractionalized excitations in quantum spin liquids has been a long- standing issue in correlated topological phases.
Conventional spectroscopic probes, such as the dynamical spin structure factor, can only detect composites of fractionalized excitations, leading to a broad continuum in energy.
Lacking a clear signature in conventional probes has been the biggest obstacle in the field. 
In this work, we theoretically investigate what kinds of distinctive signatures of fractionalized excitations can be probed in two-dimensional nonlinear spectroscopy by considering the exactly solvable Kitaev spin liquids.
We demonstrate the existence of a number of salient features of the Majorana fermions and fluxes in two-dimensional nonlinear spectroscopy, which provide crucial information about such excitations.
\end{abstract}

\maketitle

\begin{figure}[t]
\centering
\includegraphics[width=\linewidth]{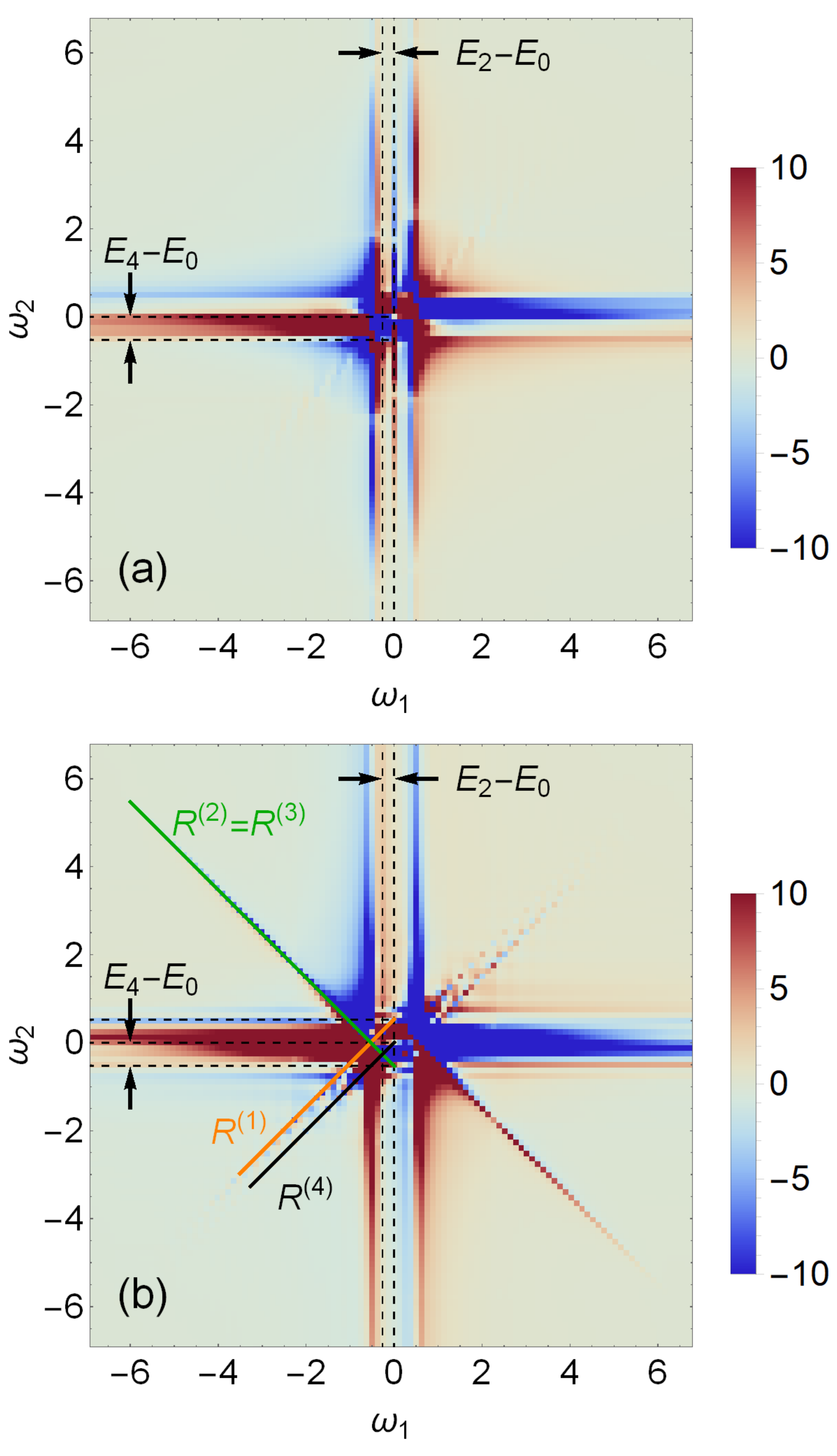}
\caption{Two-dimensional Fourier spectrum of the third-order susceptibilities (a) $\chi^{(3),z}_{zzz}(\omega_2,\omega_1,0)$ and (b) $\chi^{(3),z}_{zzz}(\omega_2,0,\omega_1)$ show the sharp vertical line signals at the two-flux gap, $\omega_1 = E_2 - E_0$. (b) $\chi^{(3),z}_{zzz}(\omega_2,0,\omega_1)$ has one sharp diagonal signal for the Majorana fermions from $R^{(3),z}_{zzz}$ and two diagonal signals from $R^{(1),z}_{zzz}$ and $R^{(4),z}_{zzz}$.
The two diagonal signals from $R^{(1),z}_{zzz}$ and $R^{(3),z}_{zzz}$ are extrapolated to finite $\omega_2$ intercept at the four-flux gap $\pm(E_4 - E_0)$, which hints at the four-flux intermediate states of the perturbative processes.}
\label{fig:2d_spectrum}
\end{figure}

Quantum spin liquids (QSLs) are prominent examples of correlated topological paramagnets that may arise due to frustrating spin interactions in Mott insulators \cite{Balents_QSLreview, Savary_QSLreview}.
The long-range quantum entanglement and ground state degeneracy, which comprises the quantum order, differentiate QSLs from trivial paramagnets and symmetry-broken phases \cite{Wen_PSG}.
Important manifestations of the quantum order are the emergent gauge fields and quasiparticles carrying fractional quantum numbers \cite{Kitaev_toric_code}.
Since the quantum entanglement is not directly observable, measuring these fractionalized excitations would be an important experimental footstep to identify quantum spin liquids.
One of the most powerful probes in magnetism, the dynamical spin structure factor measured in inelastic neutron scattering, however, shows only a broad continuum as the spin-flip involves a multitude of fractionalized excitations.
The absence of sharp signatures has hampered the progress in the discovery of quantum spin liquids.

In this Letter, we consider two-dimensional nonlinear spectroscopy as a tool to detect distinctive signatures of fractionalized quasiparticles in quantum spin liquids. 
The current work is motivated by a previous work that shows how the domain wall excitations in the transverse field Ising model can clearly be detected in two-dimensional THz spectroscopy \cite{Wan_IsingTHz}. 
Here we consider the exactly solvable Kitaev spin liquids on the honeycomb lattice \cite{Kitaev_anyon} and investigate the signatures of Majorana fermions and fluxes in two-dimensional spectroscopy.
We consider two magnetic-field pulses separated by time $\tau_1$ and measuring the nonlinear part of the induced transient magnetization at later time $\tau_2 + \tau_1$. 
The two-dimensional spectroscopy is represented by two frequencies corresponding to $\tau_1$ and $\tau_2$.
The response consists of nonlinear susceptibilities, some of which correspond to the out-of-time-order correlators of the magnetization \cite{mukamel}.
We show that the third order nonlinear susceptibilities can give rise to clear signatures of the Majorana fermions and fluxes in the Kitaev spin liquids.
We explain how one could obtain important information about such excitations from the output of the two-dimensional spectroscopy.
Our main results are shown in Fig.~\ref{fig:2d_spectrum}.

\textit{Model.}---On a honeycomb lattice with $2N$ number of sites, we consider Kitaev's spin-$\frac{1}{2}$ Hamiltonian with the isotropic strength of the bond-directional interactions \cite{Kitaev_anyon},
\begin{equation}
\hat{H} =  -  \sum_{x\text{-bond}}  \hat{\sigma}^x_{j} \hat{\sigma}^x_{k} - \sum_{y\text{-bond}} \hat{\sigma}^y_{j} \hat{\sigma}^y_{k} -  \sum_{z\text{-bond}} \hat{\sigma}^z_{j} \hat{\sigma}^z_{k},
\label{eq:model}
\end{equation}
where $\hat{\sigma}^{x,y,z}_{j}$ are the Pauli operators at site $j$.
The model has a constant of motion $\hat{W}_p = \hat{\sigma}_1^x \hat{\sigma}_2^y \hat{\sigma}_3^z \hat{\sigma}_4^x \hat{\sigma}_5^y \hat{\sigma}_6^z = \pm 1$ for each hexagonal plaquette $p$, and we say there is a static $\mathbb{Z}_2$ flux at $p$ when $\hat{W}_p = -1$.

By representing the Pauli operators in terms of Majorana fermions, $\hat{\sigma}_{j}^\alpha \doteq i b_{j}^\alpha c_{j}$, we can rewrite the model as $\mathbb{Z}_2$ gauge theory coupled to itinerant Majorana fermions,
\begin{equation}
\widetilde{H} =  \sum_{\alpha\text{-bond}} i \hat{u}^\alpha_{jk} c_{j} c_{k} \Rightarrow \widetilde{H}_u =  \sum_p \varepsilon_p \left(a^\dagger_p a_p - \frac{1}{2} \right),
\label{eq:gauge_theory}
\end{equation}
where $a_p$ represents the normal-mode complex fermions.
The model is exactly solvable because the emergent $\mathbb{Z}_2$ gauge fields $\hat{u}_{jk}^\alpha \equiv i b_j^\alpha b_k^\alpha$ commute with $\widetilde{H}$ and themselves so that the Hilbert space is factorized into the gauge (flux) sector and the matter fermion sector, $\mathcal{H} = \mathcal{H}_F \otimes \mathcal{H}_M$.
Hence, for a given gauge configuration $|F\rangle = |\{ u_{jk}^\alpha = \pm 1 \}\rangle \in \mathcal{H}_F$, $\widetilde{H}$ is reduced to a quadratic Hamiltonian $\widetilde{H}_u$ for the itinerant Majorana fermion $c$, whose eigenstates $|M\rangle$ span the matter fermion sector $\mathcal{H}_M$.

\textit{Two-dimensional spectroscopy.}---To probe the fractionalized excitations of the Kitaev spin liquid, we consider a nonlinear magnetic resonance spectroscopy with two linearly polarized, spatially uniform pulses separated by time $\tau_1$,
\begin{align}
\mathbf{B}(t) = B_0 \hat{z}  \delta(t) +  B_1 \hat{z} \delta(t-\tau_1),
\label{eq:pulse}
\end{align}
where the two incident pulses $B_0$ and $B_1$ arrive at the system at $t=0$ and $t = \tau_1$, respectively \cite{Wan_IsingTHz, Lu_magnonTHz}.
For simplicity, here we consider the case where the incident pulses are all polarized along the $\hat{z}$ direction.
These magnetic fields linearly couple to the local moments $\hat{H}_\mathrm{tot}(t) = \hat{H} - \sum_j  B^z(t) \hat{\sigma}^z_j = \hat{H} - B^z(t) \hat{M}^z$ and induce finite transient magnetization $\hat{M}^z_{01}(t)$ measured at later time $t = \tau_2 + \tau_1$.
To discard the leading contributions from the linear response, two subsequent experiments measure $\hat{M}^z_0(t)$ and $\hat{M}^z_1(t)$ due to only a single pulse $B_0$ or $B_1$, respectively. 
The nonlinear induced magnetization defined as $\hat{M}^z_\mathrm{NL}(t) = \hat{M}^z_{01}(t) -\hat{M}^z_{0}(t) - \hat{M}^z_{1}(t)$ at later time $t = \tau_1 + \tau_2$ depends only on the nonlinear dynamical responses \cite{Wan_IsingTHz},
\begin{align}
M^z_\text{NL} & (\tau_1 + \tau_2) / 2N = \chi^{(2),z}_{zz}(\tau_2,\tau_1) B_1^{z} B_0^{z} \label{eq:chi2}\\
&+ \chi^{(3),z}_{zzz}(\tau_2,\tau_1,0) B_1^{z}  B_0^{z} B_0^{z} \label{eq:chi3a} \\
&+ \chi^{(3),z}_{zzz}(\tau_2,0,\tau_1) B_1^{z}  B_1^{z} B_0^{z} + \mathcal{O}(B^4), \label{eq:chi3b}
\end{align}
where time-dependent perturbation theory gives the $n$th-order susceptibility \cite{mukamel} (we choose the unit $\hbar = 1$),
\begin{multline}
\chi^{(n),z}_{z,...,z}(\tau_n,... ,\tau_1) = \frac{i^n}{2N} \langle [[ ... [ \hat{M}^{z}(\tau_n+ ... +\tau_1), \\
\hat{M}^{z}(\tau_{n-1} + ... + \tau_1)], ... ], \hat{M}^{z}(0)]\rangle.
\label{eq:n-chi}
\end{multline}

\textit{Second-order susceptibility.}---The second-order susceptibility $\chi^{(2),z}_{zz}(\tau_2,\tau_1)$ can be calculated from the three-point correlation functions \cite{mukamel},
\begin{align}
\chi^{(2),z}_{zz}(\tau_2, \tau_1) &= \frac{i^2}{N}
\sum_{l=1}^2 \mathrm{Re} \left[Q^{(l),z}_{zz}(\tau_2,\tau_1)\right],
\end{align}
where
\begin{align}
Q^{(1),z}_{zz}(\tau_2,\tau_1) &= \langle \hat{M}^{z} (\tau_2+\tau_1) \hat{M}^{z} (\tau_1)\hat{M}^{z}(0) \rangle, \\
Q^{(2),z}_{zz}(\tau_2,\tau_1) &= - \langle \hat{M}^{z}(\tau_1) \hat{M}^{z}(\tau_2+\tau_1)\hat{M}^{z}(0) \rangle.
\end{align}

Formally, we can insert the resolution of identity $\sum_{P} |P\rangle \langle P| = \mathbbm{1}$ and decompose the three-point function into a sum of products of three matrix elements weighted by phase factors containing the dynamical information. In general,
\begin{multline}
\langle \hat{M}^z(t) \hat{M}^z(t') \hat{M}^z(0) \rangle
= \sum_{jkl}\sum_{PQ}\langle G | \hat{\sigma}_j^z | P \rangle \langle P | \hat{\sigma}_k^z |Q \rangle \\
\times \langle Q| \hat{\sigma}_l^z |G \rangle e^{i(E_G - E_P)t + i(E_P-E_Q)t'},
\label{eq:3pt_general}
\end{multline}
where $|P\rangle$ and $|Q\rangle$ are the energy eigenstates, and $|G\rangle$ is the ground state.
Since the spin operator $\hat{\sigma}_{j}^{z}$ at a $z$-bond anticommutes with $\hat{W}_p$ at the plaquettes sharing the $z$-bond, $|P\rangle$ has a pair of two adjacent fluxes.
As $\hat{\sigma}_k^z$ can either annihilate the existing two-fluxes or create two new fluxes, $|Q\rangle$ has either zero flux or two nonadjacent fluxes or four fluxes, which cannot be connected to the zero-flux state $|G\rangle$ by the single spin operator $\hat{\sigma}_l^z$.
Therefore, the second-order susceptibility should be zero under the $z$-polarized pulses.

\textit{Third-order susceptibilities.}---With the vanishing second-order susceptibility, the third-order responses determine the outcome of the nonlinear spectroscopy.
The third-order susceptibilities in Eqs.~(\ref{eq:chi3a}) and (\ref{eq:chi3b}) are calculated from the four-point correlation functions $R^{(l=1,2,3,4),z}_{zzz}$, which are expanded from the nested commutators in Eq.~(\ref{eq:n-chi}) \cite{mukamel}:
\begin{align}
\chi^{(3),z}_{zzz}(\tau_2,\tau_1,0) &= \frac{1}{N} \sum_{l=1}^4 \mathrm{Im}
\left[ R^{(l),z}_{zzz}(\tau_2,\tau_1,0) \right ], \\
\chi^{(3),z}_{zzz}(\tau_2,0, \tau_1) &= \frac{1}{N} \sum_{l=1}^4 \mathrm{Im}
\left[ R^{(l),z}_{zzz}(\tau_2,0,\tau_1) \right ],
\end{align}
where
\begin{align}
&R^{(1),z}_{zzz}(t_3,t_2,t_1) \nonumber \\
&= \langle \hat{M}^{z} (t_1) \hat{M}^{z}(t_2+t_1)\hat{M}^{z}(t_3+t_2+t_1)\hat{M}^{z}(0) \rangle,\\
&R^{(2),z}_{zzz}(t_3,t_2,t_1) \nonumber \\
&= \langle \hat{M}^{z}(0) \hat{M}^{z}(t_2+t_1)\hat{M}^{z}(t_3+t_2+t_1)\hat{M}^{z}(t_1) \rangle,\\
&R^{(3),z}_{zzz}(t_3,t_2,t_1) \nonumber \\
&= \langle \hat{M}^{z}(0) \hat{M}^{z}(t_1)\hat{M}^{z}(t_3+t_2+t_1)\hat{M}^{z}(t_2+t_1) \rangle,\\
&R^{(4),z}_{zzz}(t_3,t_2,t_1) \nonumber \\
&= \langle \hat{M}^{z} (t_3+t_2+t_1) \hat{M}^{z}(t_2+t_1)\hat{M}^{z}(t_1)\hat{M}^{z}(0) \rangle.
\end{align}

Similar to the three-point function in Eq.~(\ref{eq:3pt_general}), we can decompose the four-point functions using the resolution of identity.
For example, $R^{(3),z}_{zzz}(\tau_2, 0, \tau_1)$ becomes
\begin{align}
&R^{(3),z}_{zzz}(\tau_2, 0, \tau_1) =
\langle \hat{M}^z(0) \hat{M}^z(\tau_1) \hat{M}^z (\tau_2+ \tau_1) \hat{M}^z(\tau_1)\rangle
\nonumber \\
&= \sum_{jklm} \sum_{PQR} \langle G| \hat{\sigma}^z_j |P\rangle
\langle P| \hat{\sigma}^z_k |Q\rangle \langle Q| \hat{\sigma}^z_l |R\rangle
\langle R| \hat{\sigma}^z_m  |G \rangle \nonumber \\
&\times e^{i(E_P-E_Q)\tau_1 +i(E_Q-E_R)(\tau_2+\tau_1)+i(E_R-E_G)\tau_1}.
\label{eq:R3b} 
\end{align}
Since each spin operator flips two adjacent fluxes, $|P\rangle$ and $|R\rangle$ must belong to the two-flux sectors while $|Q\rangle$ can be either the zero-flux or nonadjacent two-flux or four-flux state.
The matrix elements for the spin operators can be calculated by rewriting $b^\alpha_j$ Majorana fermions in terms of the complex bond fermions \cite{Baskaran_Kitaev2pt, Knolle_Kitaev2pt, Knolle_Kitaev_fewPtl, Vojta_Kitaev_physical_state}. The detailed calculations can be found in the Supplemental Material \cite{SM}.

Although the above decomposition is exact, we cannot sum over an infinite number of energy eigenstates $|P\rangle$, $|Q\rangle$, $|R\rangle$. Hence, we approximate the correlation functions by truncating the summation up to intermediate states with single matter fermion \cite{Knolle_Kitaev2pt, Knolle_Kitaev_fewPtl}.
Since each spin excitation accompanies one $c$ Majorana fermion, we consider the two-flux states $|P\rangle$ and $|R\rangle$ with one matter fermion and the matter vacuum four-flux state $|Q\rangle$ \footnote{Among $N-1$ translationally inequivalent flux configurations for $|Q\rangle$, two configurations have two nonadjacent fluxes instead of four fluxes. These states contribute to the weak diagonal signals in the first and third quadrants (guided by black line in Fig.~\ref{fig:2d_spectrum}) due to $R^{(4)}$ correlations, but they do not contribute to the other diagonal signals. Our calculations found no other qualitatively distinctive features from these two-flux states compared to the four-flux state contributions.}.
This single matter fermion approximation is known to be extremely successful to calculate the dynamical spin structure factor for the Kitaev spin liquid; $97.5 \%$ of the total weight of response can be captured by the one fermion response \cite{Knolle_Kitaev2pt}.
The approximation takes advantage of the vanishing density of states of the Kitaev spin liquid at zero energy. Small perturbations would not introduce dramatic reconfiguration of the matter fermions because only few states are accessible at low energy.

\begin{figure*}[t]
\centering
\includegraphics[width=\linewidth]{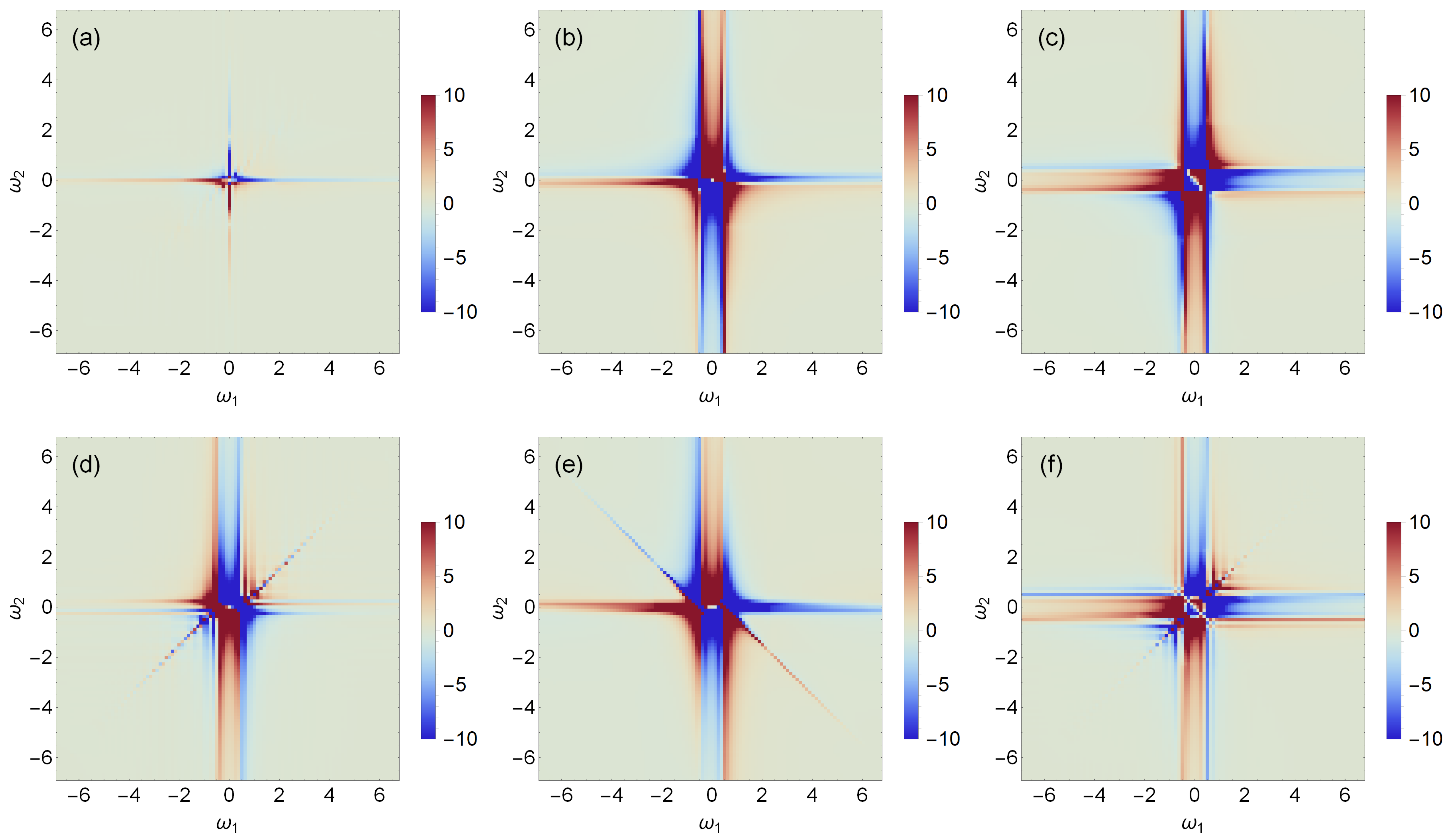}
\caption{Two-dimensional Fourier spectrum of the four point correlation functions.
Here $\mathcal{F}$ is the Fourier transformation.
\newline
(a) $\mathrm{Im}\,\mathcal{F}[\mathrm{Im}\,R^{(1,2),z}_{zzz}(\tau_2, \tau_1, 0)]$
(b) $\mathrm{Im}\,\mathcal{F}[\mathrm{Im}\,R^{(3),z}_{zzz}(\tau_2, \tau_1, 0)]$
(c) $\mathrm{Im}\,\mathcal{F}[\mathrm{Im}\,R^{(4),z}_{zzz}(\tau_2, \tau_1, 0)]$ \newline
(d) $\mathrm{Im}\,\mathcal{F}[\mathrm{Im}\,R^{(1),z}_{zzz}(\tau_2, 0, \tau_1)]$
(e) $\mathrm{Im}\,\mathcal{F}[\mathrm{Im}\,R^{(2,3),z}_{zzz}(\tau_2, 0, \tau_1)]$
(f) $\mathrm{Im}\,\mathcal{F}[\mathrm{Im}\,R^{(4),z}_{zzz}(\tau_2, 0, \tau_1)]$
}
\label{fig:correlation_function}
\end{figure*}

\textit{Results.}---We compute the real-time four-point correlation functions on a periodic lattice with $125 \times 125$ unit cells. Two-dimensional Fourier transform of the third-order susceptibilities (Fig.~\ref{fig:2d_spectrum}) and the four-point correlation functions (Fig.~\ref{fig:correlation_function})  are the main results of our work.
Here we exclude the case $|Q\rangle = |G\rangle$ in Eq.~(\ref{eq:R3b}) where the four-point function becomes nothing but a product of the two two-point functions, e.g., $\langle \hat{\sigma}^z_j \hat{\sigma}^z_k \hat{\sigma}^z_l \hat{\sigma}^z_m \rangle = \langle \hat{\sigma}^z_j \hat{\sigma}^z_k \rangle \langle \hat{\sigma}^z_l \hat{\sigma}^z_m \rangle $, which can yield physically inconsistent results within the single matter fermion approximation.

There are three distinctive features in the third-order susceptibilities in Fourier space (Fig.~\ref{fig:2d_spectrum}).
First, both $\chi^{(3),z}_{zzz}(\omega_2,\omega_1,0)$ and $\chi^{(3),z}_{zzz}(\omega_2,0,\omega_1)$, which are the Fourier transforms of $\chi^{(3),z}_{zzz}(\tau_2,\tau_1,0)$ and $\chi^{(3),z}_{zzz}(\tau_2,0,\tau_1)$ respectively, exhibit sharp vertical line signals at the two-flux gap, $\omega_1 = E_2 - E_0$. Second, $\chi^{(3),z}_{zzz}(\omega_2,0,\omega_1)$ has three extended diagonal signals. Third, if we extrapolate these diagonal signals to $\omega_1 = 0$, two of the three have an $\omega_2$ intercept equal to the four-flux gap $\pm(E_4 - E_0)$, i.e., there are overall shifts in these two diagonal signals.

While the results in Figs.~\ref{fig:2d_spectrum} and \ref{fig:correlation_function} are the direct Fourier transforms of the real-time correlation functions \cite{SM}, we can identify which processes are responsible for these distinctive signals in the susceptibilities from the formal analytic expressions of the Fourier transformed correlation functions.
For example, the Fourier transformation of $R^{(3),z}_{zzz}(\tau_2,0,\tau_1)$  [Eq.~(\ref{eq:R3b})] can be written as
\begin{align}
&R^{(3),z}_{zzz}(\omega_2,0,\omega_1) = \cdots \nonumber \\
&+ \sum_{j \neq k} \sum_{PQR} \langle G|\hat{\sigma}^z_j |P\rangle \langle P|\hat{\sigma}^z_k |Q\rangle
\langle Q|\hat{\sigma}^z_k |R\rangle \langle R| \hat{\sigma}^z_j |G\rangle \nonumber\\
&\times \frac{1}{4} \delta(\omega_1 + E_2 +\varepsilon_P -E_0)\delta(\omega_2 + E_4 - E_2 - \varepsilon_R) \nonumber \\
&+ \sum_{j\neq k} \sum_{PQR}\langle G|\hat{\sigma}^z_k |P\rangle \langle P|\hat{\sigma}^z_j |Q\rangle
\langle Q|\hat{\sigma}^z_k |R\rangle \langle R| \hat{\sigma}^z_j |G\rangle \nonumber\\
&\times \frac{1}{4} \delta(\omega_1 + E_2 +\varepsilon_P -E_0)\delta(\omega_2 + E_4 - E_2 - \varepsilon_R),
\label{eq:R3bfreq}
\end{align}
where $E_n$ is the vacuum energy of the $n$-flux state, $\varepsilon_{P(R)}$ is the matter fermion energy, and the other contributions which cannot be written in terms of the delta functions are in ($\cdots$).
The delta function pieces would show no signal for $-(E_2 - E_0) < \omega_1 < 0$. Similarly, the complex conjugate pair $R^{(3),z}_{zzz}(\omega_2,0,\omega_1)$ has no weight in $0 < \omega_1 < E_2 - E_0$. This is nothing but the well-known spin gap for spin excitations \cite{Knolle_Kitaev2pt}.
If the first pulse $B_0$ does not transfer enough energy to excite two adjacent fluxes, the Kitaev spin liquid remains in the ground state. Hence, for finite nonlinear responses, the first pulse $B_0$ must transfer energy greater than the two-flux gap $E_2 - E_0$.

Another important feature, the shifted diagonal in Fig.~\ref{fig:2d_spectrum}(b), also comes from the $R^{(3),z}_{zzz}(\omega_2,0,\omega_1)$ [Fig.~\ref{fig:correlation_function}(e)].
When $\varepsilon_P = \varepsilon_R = \varepsilon$, note that the matrix element $\langle G|\hat{\sigma}^z_j |P\rangle \langle P|\hat{\sigma}^z_k |Q\rangle
\langle Q|\hat{\sigma}^z_k |R\rangle \langle R| \hat{\sigma}^z_j |G\rangle = |\langle G|\hat{\sigma}^z_j |P\rangle \langle P|\hat{\sigma}^z_k |Q\rangle|^2 \geq 0$.
Hence, the summation over sites $\sum_{j\neq k}$, equivalently the summation over all different four-flux configurations $|Q\rangle$ excited by $\hat{\sigma}^z_j$ and $\hat{\sigma}^z_k$, results in only constructive interference. Therefore we get the strongly enhanced signal when
\begin{align}
\omega_1 &= E_0 - E_2 - \varepsilon < 0, \label{eq:fluxGap2} \\
\omega_2 &= E_2 - E_4 + \varepsilon = - \omega_1 - (E_4 - E_0), \label{eq:yintercept}
\end{align}
which corresponds to the shifted diagonal with the slope of $-1$ and the $\omega_2$ intercept $-(E_4 - E_0)$. According to Eq.~(\ref{eq:fluxGap2}), the domain of the line is determined by the single matter fermion bandwidth and the two-flux gap, and this is confirmed by Figs. \ref{fig:2d_spectrum}~(b) and \ref{fig:correlation_function}~(e).

Following a similar logic, we can understand two-flux gaps and the other two coherent diagonal signals coming from $R^{(1),z}_{zzz}(\omega_2, 0, \omega_1)$ and $R^{(4),z}_{zzz}(\omega_2, 0, \omega_1)$, which have contributions with the constraints in the sum over intermediate states via $\delta(\omega_1 + E_0 - E_2 - \varepsilon_R)\delta(\omega_2 + E_4 - E_2 - \varepsilon_R)$ and $\delta(\omega_1 + E_0 -E_2 - \varepsilon_R)\delta(\omega_2 + E_0 - E_2 -\varepsilon_P)$, respectively.
$R^{(1),z}_{zzz}(\omega_2, 0, \omega_1)$ yields the shifted diagonal $\omega_2 = \omega_1 - (E_4 - E_0)$, and $R^{(4),z}_{zzz}(\omega_2, 0, \omega_1)$ gives
$\omega_2 = \omega_1$ from the constructive interference with $\varepsilon_P = \varepsilon_R$ for $\omega_1 \geq E_2 - E_0$.

\textit{Conclusion.}---In this work, we have demonstrated how two-dimensional spectroscopy can be used to obtain useful information about fractionalized excitations in the Kitaev spin liquids, where the single spin-flip process excites a Majorana fermion and two fluxes in adjacent plaquettes.
The spectroscopic signatures as a function of two frequencies, $\omega_1$ and $\omega_2$, corresponding to the delay time of two successive magnetic pulses and the time of measurement, offer a clear identification of both the Majorana fermions and flux excitations.
We demonstrated that the two-flux gap appears in $\omega_1$ and the shifted diagonal signal in the $\omega_1$-$\omega_2$ plane has an $\omega_2$ intercept at the four-flux gap.
Most importantly, the presence of the sharp diagonal signals is the direct consequence of the itinerant Majorana fermions.
The domain of finite response in the two-frequency $\omega_1$-$\omega_2$ plane is determined by a number of stringent conditions, which makes it possible to identify clear signatures of fractionalized excitations.
It will be interesting to extend our work to other theoretical models of quantum spin liquids, that are not exactly solvable.
Furthermore, the results reported here may be tested in a number of candidate materials for the Kitaev spin liquids \cite{JKmechanism, Trebst_Kitaev,William_review, Winter_review, Rau_review, Takagi_KitaevReview, Motome_MajoranaHunting}.
We expect our results will shed significant light on the identification of fractionalized excitations and quantum spin liquids.

This work was supported by the NSERC of Canada and the Center for Quantum Materials at the University of Toronto. Y.B.K. is supported by the Killam Research Fellowship of the Canada Council for the Arts, and K.H.L is supported by the Institute for Basic Science (IBS) in Korea (IBS-R009-G1).
We thank Center for Advanced Computation at Korea Institute for Advanced Study for providing computing resources for this work.
We thank Y. Wan and N. P. Armitage for letting us know about their work.
We also thank Masafumi Udagawa, Qiang-Hua Wang, Panagiotis P. Stavropoulos and Adarsh S. Patri for helpful discussion.

W.C. and K.H.L. contributed equally to this work.  

\bibliography{KitaevTHz}

\begin{thebibliography}{23}%
\makeatletter
\providecommand \@ifxundefined [1]{%
 \@ifx{#1\undefined}
}%
\providecommand \@ifnum [1]{%
 \ifnum #1\expandafter \@firstoftwo
 \else \expandafter \@secondoftwo
 \fi
}%
\providecommand \@ifx [1]{%
 \ifx #1\expandafter \@firstoftwo
 \else \expandafter \@secondoftwo
 \fi
}%
\providecommand \natexlab [1]{#1}%
\providecommand \enquote  [1]{``#1''}%
\providecommand \bibnamefont  [1]{#1}%
\providecommand \bibfnamefont [1]{#1}%
\providecommand \citenamefont [1]{#1}%
\providecommand \href@noop [0]{\@secondoftwo}%
\providecommand \href [0]{\begingroup \@sanitize@url \@href}%
\providecommand \@href[1]{\@@startlink{#1}\@@href}%
\providecommand \@@href[1]{\endgroup#1\@@endlink}%
\providecommand \@sanitize@url [0]{\catcode `\\12\catcode `\$12\catcode
  `\&12\catcode `\#12\catcode `\^12\catcode `\_12\catcode `\%12\relax}%
\providecommand \@@startlink[1]{}%
\providecommand \@@endlink[0]{}%
\providecommand \url  [0]{\begingroup\@sanitize@url \@url }%
\providecommand \@url [1]{\endgroup\@href {#1}{\urlprefix }}%
\providecommand \urlprefix  [0]{URL }%
\providecommand \Eprint [0]{\href }%
\providecommand \doibase [0]{http://dx.doi.org/}%
\providecommand \selectlanguage [0]{\@gobble}%
\providecommand \bibinfo  [0]{\@secondoftwo}%
\providecommand \bibfield  [0]{\@secondoftwo}%
\providecommand \translation [1]{[#1]}%
\providecommand \BibitemOpen [0]{}%
\providecommand \bibitemStop [0]{}%
\providecommand \bibitemNoStop [0]{.\EOS\space}%
\providecommand \EOS [0]{\spacefactor3000\relax}%
\providecommand \BibitemShut  [1]{\csname bibitem#1\endcsname}%
\let\auto@bib@innerbib\@empty
\bibitem [{\citenamefont {Balents}(2010)}]{Balents_QSLreview}%
  \BibitemOpen
  \bibfield  {author} {\bibinfo {author} {\bibfnamefont {L.}~\bibnamefont
  {Balents}},\ }\href {https://doi.org/10.1038/nature08917} {\bibfield
  {journal} {\bibinfo  {journal} {Nature}\ }\textbf {\bibinfo {volume} {464}},\
  \bibinfo {pages} {199} (\bibinfo {year} {2010})}\BibitemShut {NoStop}%
\bibitem [{\citenamefont {Savary}\ and\ \citenamefont
  {Balents}(2017)}]{Savary_QSLreview}%
  \BibitemOpen
  \bibfield  {author} {\bibinfo {author} {\bibfnamefont {L.}~\bibnamefont
  {Savary}}\ and\ \bibinfo {author} {\bibfnamefont {L.}~\bibnamefont
  {Balents}},\ }\href {\doibase 10.1088/0034-4885/80/1/016502} {\bibfield
  {journal} {\bibinfo  {journal} {Reports on Progress in Physics}\ }\textbf
  {\bibinfo {volume} {80}},\ \bibinfo {pages} {016502} (\bibinfo {year}
  {2017})}\BibitemShut {NoStop}%
\bibitem [{\citenamefont {Wen}(2002)}]{Wen_PSG}%
  \BibitemOpen
  \bibfield  {author} {\bibinfo {author} {\bibfnamefont {X.-G.}\ \bibnamefont
  {Wen}},\ }\href {\doibase 10.1103/PhysRevB.65.165113} {\bibfield  {journal}
  {\bibinfo  {journal} {Phys. Rev. B}\ }\textbf {\bibinfo {volume} {65}},\
  \bibinfo {pages} {165113} (\bibinfo {year} {2002})}\BibitemShut {NoStop}%
\bibitem [{\citenamefont {Kitaev}(2003)}]{Kitaev_toric_code}%
  \BibitemOpen
  \bibfield  {author} {\bibinfo {author} {\bibfnamefont {A.}~\bibnamefont
  {Kitaev}},\ }\href {\doibase https://doi.org/10.1016/S0003-4916(02)00018-0}
  {\bibfield  {journal} {\bibinfo  {journal} {Annals of Physics}\ }\textbf
  {\bibinfo {volume} {303}},\ \bibinfo {pages} {2 } (\bibinfo {year}
  {2003})}\BibitemShut {NoStop}%
\bibitem [{\citenamefont {Wan}\ and\ \citenamefont
  {Armitage}(2019)}]{Wan_IsingTHz}%
  \BibitemOpen
  \bibfield  {author} {\bibinfo {author} {\bibfnamefont {Y.}~\bibnamefont
  {Wan}}\ and\ \bibinfo {author} {\bibfnamefont {N.~P.}\ \bibnamefont
  {Armitage}},\ }\href {\doibase 10.1103/PhysRevLett.122.257401} {\bibfield
  {journal} {\bibinfo  {journal} {Phys. Rev. Lett.}\ }\textbf {\bibinfo
  {volume} {122}},\ \bibinfo {pages} {257401} (\bibinfo {year}
  {2019})}\BibitemShut {NoStop}%
\bibitem [{\citenamefont {Kitaev}(2006)}]{Kitaev_anyon}%
  \BibitemOpen
  \bibfield  {author} {\bibinfo {author} {\bibfnamefont {A.}~\bibnamefont
  {Kitaev}},\ }\href {\doibase 10.1016/j.aop.2005.10.005} {\bibfield  {journal}
  {\bibinfo  {journal} {Ann. Phys.}\ }\textbf {\bibinfo {volume} {321}},\
  \bibinfo {pages} {2} (\bibinfo {year} {2006})}\BibitemShut {NoStop}%
\bibitem [{\citenamefont {Mukamel}(1995)}]{mukamel}%
  \BibitemOpen
  \bibfield  {author} {\bibinfo {author} {\bibfnamefont {S.}~\bibnamefont
  {Mukamel}},\ }\href@noop {} {\emph {\bibinfo {title} {{Principles of
  Nonlinear Optical Spectroscopy}}}}\ (\bibinfo  {publisher} {Oxford University
  Press},\ \bibinfo {year} {1995})\BibitemShut {NoStop}%
\bibitem [{\citenamefont {Lu}\ \emph {et~al.}(2017)\citenamefont {Lu},
  \citenamefont {Li}, \citenamefont {Hwang}, \citenamefont {Ofori-Okai},
  \citenamefont {Kurihara}, \citenamefont {Suemoto},\ and\ \citenamefont
  {Nelson}}]{Lu_magnonTHz}%
  \BibitemOpen
  \bibfield  {author} {\bibinfo {author} {\bibfnamefont {J.}~\bibnamefont
  {Lu}}, \bibinfo {author} {\bibfnamefont {X.}~\bibnamefont {Li}}, \bibinfo
  {author} {\bibfnamefont {H.~Y.}\ \bibnamefont {Hwang}}, \bibinfo {author}
  {\bibfnamefont {B.~K.}\ \bibnamefont {Ofori-Okai}}, \bibinfo {author}
  {\bibfnamefont {T.}~\bibnamefont {Kurihara}}, \bibinfo {author}
  {\bibfnamefont {T.}~\bibnamefont {Suemoto}}, \ and\ \bibinfo {author}
  {\bibfnamefont {K.~A.}\ \bibnamefont {Nelson}},\ }\href {\doibase
  10.1103/PhysRevLett.118.207204} {\bibfield  {journal} {\bibinfo  {journal}
  {Phys. Rev. Lett.}\ }\textbf {\bibinfo {volume} {118}},\ \bibinfo {pages}
  {207204} (\bibinfo {year} {2017})}\BibitemShut {NoStop}%
\bibitem [{\citenamefont {Baskaran}\ \emph {et~al.}(2007)\citenamefont
  {Baskaran}, \citenamefont {Mandal},\ and\ \citenamefont
  {Shankar}}]{Baskaran_Kitaev2pt}%
  \BibitemOpen
  \bibfield  {author} {\bibinfo {author} {\bibfnamefont {G.}~\bibnamefont
  {Baskaran}}, \bibinfo {author} {\bibfnamefont {S.}~\bibnamefont {Mandal}}, \
  and\ \bibinfo {author} {\bibfnamefont {R.}~\bibnamefont {Shankar}},\ }\href
  {\doibase 10.1103/PhysRevLett.98.247201} {\bibfield  {journal} {\bibinfo
  {journal} {Phys. Rev. Lett.}\ }\textbf {\bibinfo {volume} {98}},\ \bibinfo
  {pages} {247201} (\bibinfo {year} {2007})}\BibitemShut {NoStop}%
\bibitem [{\citenamefont {Knolle}\ \emph {et~al.}(2014)\citenamefont {Knolle},
  \citenamefont {Kovrizhin}, \citenamefont {Chalker},\ and\ \citenamefont
  {Moessner}}]{Knolle_Kitaev2pt}%
  \BibitemOpen
  \bibfield  {author} {\bibinfo {author} {\bibfnamefont {J.}~\bibnamefont
  {Knolle}}, \bibinfo {author} {\bibfnamefont {D.~L.}\ \bibnamefont
  {Kovrizhin}}, \bibinfo {author} {\bibfnamefont {J.~T.}\ \bibnamefont
  {Chalker}}, \ and\ \bibinfo {author} {\bibfnamefont {R.}~\bibnamefont
  {Moessner}},\ }\href {\doibase 10.1103/PhysRevLett.112.207203} {\bibfield
  {journal} {\bibinfo  {journal} {Phys. Rev. Lett.}\ }\textbf {\bibinfo
  {volume} {112}},\ \bibinfo {pages} {207203} (\bibinfo {year}
  {2014})}\BibitemShut {NoStop}%
\bibitem [{\citenamefont {Knolle}\ \emph {et~al.}(2015)\citenamefont {Knolle},
  \citenamefont {Kovrizhin}, \citenamefont {Chalker},\ and\ \citenamefont
  {Moessner}}]{Knolle_Kitaev_fewPtl}%
  \BibitemOpen
  \bibfield  {author} {\bibinfo {author} {\bibfnamefont {J.}~\bibnamefont
  {Knolle}}, \bibinfo {author} {\bibfnamefont {D.~L.}\ \bibnamefont
  {Kovrizhin}}, \bibinfo {author} {\bibfnamefont {J.~T.}\ \bibnamefont
  {Chalker}}, \ and\ \bibinfo {author} {\bibfnamefont {R.}~\bibnamefont
  {Moessner}},\ }\href {\doibase 10.1103/PhysRevB.92.115127} {\bibfield
  {journal} {\bibinfo  {journal} {Phys. Rev. B}\ }\textbf {\bibinfo {volume}
  {92}},\ \bibinfo {pages} {115127} (\bibinfo {year} {2015})}\BibitemShut
  {NoStop}%
\bibitem [{\citenamefont {Zschocke}\ and\ \citenamefont
  {Vojta}(2015)}]{Vojta_Kitaev_physical_state}%
  \BibitemOpen
  \bibfield  {author} {\bibinfo {author} {\bibfnamefont {F.}~\bibnamefont
  {Zschocke}}\ and\ \bibinfo {author} {\bibfnamefont {M.}~\bibnamefont
  {Vojta}},\ }\href {\doibase 10.1103/PhysRevB.92.014403} {\bibfield  {journal}
  {\bibinfo  {journal} {Phys. Rev. B}\ }\textbf {\bibinfo {volume} {92}},\
  \bibinfo {pages} {014403} (\bibinfo {year} {2015})}\BibitemShut {NoStop}%
\bibitem [{SM()}]{SM}%
  \BibitemOpen
  \href@noop {} {\ }\bibinfo {note} {See Supplemental Materials for detailed
  discussions.}\BibitemShut {Stop}%
\bibitem [{Note1()}]{Note1}%
  \BibitemOpen
  \bibinfo {note} {Among $N-1$ translationally inequivalent flux configurations
  for $|Q\rangle $, two configurations have two nonadjacent fluxes instead of
  four fluxes. These states contribute to the weak diagonal signals in the
  first and third quadrants (guided by black line in Fig.~\ref
  {fig:2d_spectrum}) due to $R^{(4)}$ correlations, but they do not contribute
  to the other diagonal signals. Our calculations found no other qualitatively
  distinctive features from these two-flux states compared to the four-flux
  state contributions.}\BibitemShut {Stop}%
\bibitem [{\citenamefont {Jackeli}\ and\ \citenamefont
  {Khaliullin}(2009)}]{JKmechanism}%
  \BibitemOpen
  \bibfield  {author} {\bibinfo {author} {\bibfnamefont {G.}~\bibnamefont
  {Jackeli}}\ and\ \bibinfo {author} {\bibfnamefont {G.}~\bibnamefont
  {Khaliullin}},\ }\href {\doibase 10.1103/PhysRevLett.102.017205} {\bibfield
  {journal} {\bibinfo  {journal} {Phys. Rev. Lett.}\ }\textbf {\bibinfo
  {volume} {102}},\ \bibinfo {pages} {017205} (\bibinfo {year}
  {2009})}\BibitemShut {NoStop}%
\bibitem [{\citenamefont {{Trebst}}()}]{Trebst_Kitaev}%
  \BibitemOpen
  \bibfield  {author} {\bibinfo {author} {\bibfnamefont {S.}~\bibnamefont
  {{Trebst}}},\ }\href {https://arxiv.org/abs/1701.07056} {\enquote {\bibinfo
  {title} {Kitaev materials},}\ }\Eprint {http://arxiv.org/abs/1701.07056}
  {arXiv:1701.07056 [cond-mat.str-el]} \BibitemShut {NoStop}%
\bibitem [{\citenamefont {Witczak-Krempa}\ \emph {et~al.}(2014)\citenamefont
  {Witczak-Krempa}, \citenamefont {Chen}, \citenamefont {Kim},\ and\
  \citenamefont {Balents}}]{William_review}%
  \BibitemOpen
  \bibfield  {author} {\bibinfo {author} {\bibfnamefont {W.}~\bibnamefont
  {Witczak-Krempa}}, \bibinfo {author} {\bibfnamefont {G.}~\bibnamefont
  {Chen}}, \bibinfo {author} {\bibfnamefont {Y.~B.}\ \bibnamefont {Kim}}, \
  and\ \bibinfo {author} {\bibfnamefont {L.}~\bibnamefont {Balents}},\ }\href
  {\doibase 10.1146/annurev-conmatphys-020911-125138} {\bibfield  {journal}
  {\bibinfo  {journal} {Annu. Rev. Condens. Matter Phys.}\ }\textbf {\bibinfo
  {volume} {5}},\ \bibinfo {pages} {57} (\bibinfo {year} {2014})}\BibitemShut
  {NoStop}%
\bibitem [{\citenamefont {Winter}\ \emph {et~al.}(2017)\citenamefont {Winter},
  \citenamefont {Tsirlin}, \citenamefont {Daghofer}, \citenamefont {van~den
  Brink}, \citenamefont {Singh}, \citenamefont {Gegenwart},\ and\ \citenamefont
  {Valent{\'{\i}}}}]{Winter_review}%
  \BibitemOpen
  \bibfield  {author} {\bibinfo {author} {\bibfnamefont {S.~M.}\ \bibnamefont
  {Winter}}, \bibinfo {author} {\bibfnamefont {A.~A.}\ \bibnamefont {Tsirlin}},
  \bibinfo {author} {\bibfnamefont {M.}~\bibnamefont {Daghofer}}, \bibinfo
  {author} {\bibfnamefont {J.}~\bibnamefont {van~den Brink}}, \bibinfo {author}
  {\bibfnamefont {Y.}~\bibnamefont {Singh}}, \bibinfo {author} {\bibfnamefont
  {P.}~\bibnamefont {Gegenwart}}, \ and\ \bibinfo {author} {\bibfnamefont
  {R.}~\bibnamefont {Valent{\'{\i}}}},\ }\href {\doibase
  10.1088/1361-648x/aa8cf5} {\bibfield  {journal} {\bibinfo  {journal} {Journal
  of Physics: Condensed Matter}\ }\textbf {\bibinfo {volume} {29}},\ \bibinfo
  {pages} {493002} (\bibinfo {year} {2017})}\BibitemShut {NoStop}%
\bibitem [{\citenamefont {Rau}\ \emph {et~al.}(2016)\citenamefont {Rau},
  \citenamefont {Lee},\ and\ \citenamefont {Kee}}]{Rau_review}%
  \BibitemOpen
  \bibfield  {author} {\bibinfo {author} {\bibfnamefont {J.~G.}\ \bibnamefont
  {Rau}}, \bibinfo {author} {\bibfnamefont {E.~K.-H.}\ \bibnamefont {Lee}}, \
  and\ \bibinfo {author} {\bibfnamefont {H.-Y.}\ \bibnamefont {Kee}},\ }\href
  {\doibase 10.1146/annurev-conmatphys-031115-011319} {\bibfield  {journal}
  {\bibinfo  {journal} {Annu. Rev. Condens. Matter Phys.}\ }\textbf {\bibinfo
  {volume} {7}},\ \bibinfo {pages} {195} (\bibinfo {year} {2016})}\BibitemShut
  {NoStop}%
\bibitem [{\citenamefont {Takagi}\ \emph {et~al.}(2019)\citenamefont {Takagi},
  \citenamefont {Takayama}, \citenamefont {Jackeli}, \citenamefont
  {Khaliullin},\ and\ \citenamefont {Nagler}}]{Takagi_KitaevReview}%
  \BibitemOpen
  \bibfield  {author} {\bibinfo {author} {\bibfnamefont {H.}~\bibnamefont
  {Takagi}}, \bibinfo {author} {\bibfnamefont {T.}~\bibnamefont {Takayama}},
  \bibinfo {author} {\bibfnamefont {G.}~\bibnamefont {Jackeli}}, \bibinfo
  {author} {\bibfnamefont {G.}~\bibnamefont {Khaliullin}}, \ and\ \bibinfo
  {author} {\bibfnamefont {S.~E.}\ \bibnamefont {Nagler}},\ }\href {\doibase
  10.1038/s42254-019-0038-2} {\bibfield  {journal} {\bibinfo  {journal} {Nature
  Reviews Physics}\ }\textbf {\bibinfo {volume} {1}},\ \bibinfo {pages} {264}
  (\bibinfo {year} {2019})}\BibitemShut {NoStop}%
\bibitem [{\citenamefont {Motome}\ and\ \citenamefont
  {Nasu}(2020)}]{Motome_MajoranaHunting}%
  \BibitemOpen
  \bibfield  {author} {\bibinfo {author} {\bibfnamefont {Y.}~\bibnamefont
  {Motome}}\ and\ \bibinfo {author} {\bibfnamefont {J.}~\bibnamefont {Nasu}},\
  }\href {\doibase 10.7566/JPSJ.89.012002} {\bibfield  {journal} {\bibinfo
  {journal} {Journal of the Physical Society of Japan}\ }\textbf {\bibinfo
  {volume} {89}},\ \bibinfo {pages} {012002} (\bibinfo {year}
  {2020})}\BibitemShut {NoStop}%
\bibitem [{\citenamefont {Pedrocchi}\ \emph {et~al.}(2011)\citenamefont
  {Pedrocchi}, \citenamefont {Chesi},\ and\ \citenamefont
  {Loss}}]{Pedrocchi_Kitaev_physical_state}%
  \BibitemOpen
  \bibfield  {author} {\bibinfo {author} {\bibfnamefont {F.~L.}\ \bibnamefont
  {Pedrocchi}}, \bibinfo {author} {\bibfnamefont {S.}~\bibnamefont {Chesi}}, \
  and\ \bibinfo {author} {\bibfnamefont {D.}~\bibnamefont {Loss}},\ }\href
  {\doibase 10.1103/PhysRevB.84.165414} {\bibfield  {journal} {\bibinfo
  {journal} {Phys. Rev. B}\ }\textbf {\bibinfo {volume} {84}},\ \bibinfo
  {pages} {165414} (\bibinfo {year} {2011})}\BibitemShut {NoStop}%
\bibitem [{\citenamefont {Blaizot}\ and\ \citenamefont
  {Ripka}(1986)}]{Blaizot_quantum_theory}%
  \BibitemOpen
  \bibfield  {author} {\bibinfo {author} {\bibfnamefont {J.-P.}\ \bibnamefont
  {Blaizot}}\ and\ \bibinfo {author} {\bibfnamefont {G.}~\bibnamefont
  {Ripka}},\ }\href@noop {} {\emph {\bibinfo {title} {{Quantum Theory of Finite
  Systems}}}}\ (\bibinfo  {publisher} {MIT Press},\ \bibinfo {year}
  {1986})\BibitemShut {NoStop}%
\end{thebibliography}%

\clearpage
\onecolumngrid
\setcounter{equation}{0}
\setcounter{figure}{0}
\setcounter{table}{0}
\setcounter{page}{1}
\setcounter{section}{0}
\setcounter{subsection}{0}

\pagebreak
\widetext
\begin{center}
\textbf{\large Supplemental Materials for ``Theory of Two-Dimensional Nonlinear Spectroscopy for the Kitaev Spin Liquid"}
\end{center}
\setcounter{equation}{0}
\setcounter{figure}{0}
\setcounter{table}{0}
\setcounter{page}{1}
\makeatletter
\renewcommand{\theequation}{S\arabic{equation}}
\renewcommand{\thefigure}{S\arabic{figure}}

\section{Nonlinear susceptibilities from correlation functions}

Two-dimensional spectroscopy measures nonlinear dynamical susceptibilities of correlated systems. In this section, we briefly review nonlinear response theory and derive the nonlinear susceptibilities from multi-spin correlation functions. More interested readers may refer to Ref.~\cite{mukamel} and references therein. Here, we choose the unit $\hbar = 1$.

Consider a Hamiltonian $\hat{H}_T (t)= \hat{H} + \hat{V}(t)$ with a weak light-matter interaction $\hat{V}(t) = - \mathbf{B}(t) \cdot \hat{\mathbf{M}}$.
By treating the time-dependent potential $\hat{V}(t)$ as a small perturbation, we solve the Liouville equation,
$i \partial_t \hat{\rho}= [ \hat{H}_T, \hat{\rho} ]$,
to get the magnetization induced by the incoming magnetic fields.
Because of the structural similarity between the Liouville equation and the Schr\"{o}dinger equation ($i \partial_t \psi = \hat{H}_T \psi$),
the Liouville equation can be considered as the ``Schr\"{o}dinger equation" $i\partial_t \hat{\rho} =L_T \hat{\rho}$ in a vector space of density operators, so called the Liouville space, with the superoperator $L_T: \hat{\rho} \mapsto [\hat{H}_T, \hat{\rho}]$.
Hence, with the analogy $\hat{H}_T \leftrightarrow L_T$, we can adopt known time-dependent perturbation theory for wavefunctions in the Hilbert space to solve for density operators in the Liouville space.

In the interaction picture, the operator evolves as $\hat{\mathcal{O}}_I(t) = e^{i\hat{H}(t-t_0)} \hat{\mathcal{O}} e^{-i\hat{H}(t-t_0)}$ while the density operator obeys
\begin{align}
i \frac{\partial \hat{\rho}_I}{\partial t} = [ \hat{V}_I(t) , \hat{\rho}_I(t) ].
\end{align}
So the Dyson series in the Liouville space is $\hat{\rho}_I(t) = \sum_{n=0}^{\infty} \hat{\rho}_I^{(n)}(t)$, where
\begin{align}
&\hat{\rho}_I^{(0)}(t) = \hat{\rho}(t_0) = |G\rangle \langle G|, \\
&\hat{\rho}_I^{(1)}(t) = - i \int_{t_0}^t [\hat{V}_I(t_1),\hat{\rho}(t_0) ] dt_1,\\
&\hat{\rho}_I^{(n>1)}(t) = \left( - i \right)^n \int_{t_0}^t \int_{t_0}^{t_n} \cdots \int_{t_0}^{t_2} [ \hat{V}_I(t_n), [\hat{V}_I(t_{n-1}), [\cdots,[\hat{V}_I(t_1), \hat{\rho}(t_0) ] \cdots ] ] ] dt_1 \cdots dt_n
\end{align}
if we prepared the ground state of $\hat{H}$, $\hat{\rho}(t_0) = |G\rangle \langle G|$, at time $t=t_0$.
Then the induced magnetization is $M^\alpha(t) = \sum_{n=0}^\infty M^{(n),\alpha}(t)$, where the $n$th order contribution to the induced magnetization is
\begin{align}
&M^{(n),\alpha} = \mathrm{Tr}\left( \hat{M}^\alpha_I(t) \hat{\rho}_I^{(n)}(t) \right) \\
&=i^n \int_{t_0}^t \int_{t_0}^{t_n} \cdots \int_{t_0}^{t_2} \mathrm{Tr}
\left( \hat{M}^\alpha_I(t) [\hat{M}^{\alpha_n}_I(t_n), [ \cdots ,[ \hat{M}^{\alpha_1}_I(t_1), \hat{\rho}(t_0)] \cdots ] ] \right) B^{\alpha_n}(t_n) \cdots B^{\alpha_1}(t_1) ~dt_1 \cdots dt_n \\
&= i^n \int_{t_0}^t \int_{t_0}^{t_n} \cdots \int_{t_0}^{t_2} \mathrm{Tr}
\left( [ [\cdots [\hat{M}^\alpha_I(t), \hat{M}^{\alpha_n}_I(t_n)], \cdots] ,\hat{M}^{\alpha_1}_I(t_1)]\hat{\rho}(t_0) \right) B^{\alpha_n}(t_n) \cdots B^{\alpha_1}(t_1) ~dt_1 \cdots dt_n.
\end{align}
To simplify the expression, let us introduce new variables $\tau_m = t_{m+1} - t_m$, $t_{n+1} \equiv t$ and take the limit $t_0 \to -\infty$. Then we can deduce the nonlinear magnetization from the multi-spin correlation functions:
\begin{align}
M^{(n),\alpha}(t) = N_\text{site} \int_0^\infty \cdots \int_0^\infty \chi^{(n),\alpha}_{\alpha_n,...,\alpha_1}(\tau_n,...,\tau_1) B^{\alpha_n}(t-\tau_n) \cdots B^{\alpha_1}(t-\tau_n-\tau_{n-1} -...- \tau_1),
\end{align}
with the $n$th order nonlinear susceptibility
\begin{align}
\chi^{(n),\alpha}_{\alpha_n,...,\alpha_1}(\tau_n,...,\tau_1) &= \frac{i^n}{N_\text{site}} \theta(\tau_n) \cdots \theta(\tau_1)
\mathrm{Tr} \left( [ [\cdots [\hat{M}^\alpha_I(\tau_n + ... + \tau_1), \hat{M}^{\alpha_n}_I(\tau_{n-1} + ... +\tau_1)], \cdots] ,\hat{M}^{\alpha_1}_I(0)]\hat{\rho}(-\infty) \right) \nonumber \\
&=\frac{i^n}{N_\text{site}} \theta(\tau_n) \cdots \theta(\tau_1)  \langle G |  [ [\cdots [\hat{M}^\alpha_I(\tau_n +...+\tau_1), \hat{M}^{\alpha_n}_I(\tau_{n-1} + ... + \tau_1)], \cdots] ,\hat{M}^{\alpha_1}_I(0)]  | G \rangle,
\label{eq:suscep}
\end{align}
where $\theta(x)$ is the Heaviside step function imposing the causality.

\section{Physical Hilbert space of the Kitaev model}

The Kitaev model is solved exactly by the Majorana fermion representation of the spin operators, $\hat{\sigma}_{j}^\alpha = i b_{j}^\alpha c_{j}$,
\begin{align}
\hat{H} = - \sum_{\alpha\text{-bond}} \sigma_j^\alpha \sigma_k^\alpha \Rightarrow
\widetilde{H} = \sum_{\alpha\text{-bond}} (i b_j^\alpha b_k^\alpha) (ic_j c_k)
\equiv  \sum_{\alpha\text{-bond}} i \hat{u}^\alpha_{jk} c_j c_k
\end{align}
where
$\{ b_{j}^\alpha, b_{k}^\beta \} = 2 \delta_{jk}\delta^{\alpha\beta}$, $\{ b_{j}^\alpha, c_{k} \} = 0$, and $\{c_{j}, c_{k} \} = 2\delta_{jk}$.
Since the $\mathbb{Z}_2$ gauge fields $\hat{u}^\alpha_{jk}$ commute with $\widetilde{H}$ and themselves, the fermionic Hilbert space $\mathcal{H} = \mathcal{H}_F \otimes \mathcal{H}_M$ is factorized into the gauge or flux sector $\mathcal{H}_F$ and the matter fermion sector $\mathcal{H}_M$.

To represent the gauge configurations $\{\hat{u}_{jk}^\alpha\}$, we introduce a complex bond fermion at each bond
\begin{align}
\chi_{\langle jk \rangle}^\alpha = \frac{1}{2} \left( b_j^\alpha + i b_k^\alpha \right),~ \chi_{\langle jk \rangle}^{\alpha\dagger} = \frac{1}{2} \left( b_j^\alpha - i b_k^\alpha \right),
\end{align}
where site $j$, $k$ belongs to the sublattice $s = A$ and $s = B$, respectively \cite{Baskaran_Kitaev2pt}. 
Then $\hat{u}_{jk}^\alpha = 2 \chi_{\langle jk\rangle_\alpha}^{\dagger}\chi_{\langle jk \rangle_\alpha} - 1$.
Therefore, the Hilbert space for the $\mathbb{Z}_2$ gauge fields $\hat{u}_{jk}^\alpha$ is equivalent to the $2^{3N}$-dimensional fermionic Hilbert space $\mathcal{H}_F = \mathrm{span} \left \{ | F \rangle  \right \}$ for $\chi^\alpha_{\langle jk \rangle}$, where
\begin{align}
\left | F \right \rangle = \prod_{{\langle jk \rangle}}
\left( \chi^\alpha_{\langle jk \rangle} \right)^{n_{\langle jk \rangle}} \left |F_0 \right\rangle
\end{align}
with $n_{{\langle jk \rangle}_\alpha} = 0, 1$ and the zero-flux sector $|F_0\rangle$ such that
$\chi^{\alpha\dagger}_{\langle jk \rangle} | F_0 \rangle = 0$ for all bonds $\langle jk \rangle$.

After we fix the gauge for a given flux sector, $\widetilde{H}$ becomes an exactly solvable free fermion Hamiltonian
\begin{align}
\widetilde{H}_u &= \frac{i}{2} \sum_{\mu,\nu} M_{\mu\nu}c_{\mu,A} c_{\nu,B}
= \frac{i}{4}
\begin{bmatrix}
\mathbf{c}_{A}^T & \mathbf{c}_{B}^T
\end{bmatrix}
\begin{bmatrix}
0 & M \\
-M^T & 0
\end{bmatrix}
\begin{bmatrix}
\mathbf{c}_{A} \\
\mathbf{c}_{B}
\end{bmatrix},
\end{align}
where a real matrix $M_{\mu\nu} = 2 u_{\mu\nu}^\alpha$ for the $\alpha$-bond and zero otherwise, and $\mu,\nu=1,...,N$ label the unit cell, i.e., $j=(\mu,s)$.
$\widetilde{H}_u$ can be block-diagonalized by a special orthogonal transformation $Q \in O(2N)$ \cite{Kitaev_anyon, Vojta_Kitaev_physical_state}. To be specific,
\begin{align}
\widetilde{H}_u &=
\frac{i}{4}
\begin{bmatrix}
\mathbf{c}_{A}^T & \mathbf{c}_{B}^T
\end{bmatrix}
\begin{bmatrix}
0 & M \\
-M^T & 0
\end{bmatrix}
\begin{bmatrix}
\mathbf{c}_{A} \\
\mathbf{c}_{B}
\end{bmatrix}
= \frac{i}{4}
\begin{bmatrix}
\mathbf{c}_{A}^T & \mathbf{c}_{B}^T
\end{bmatrix}
\begin{bmatrix}
0 & U S V^T \\
-V S U^T & 0
\end{bmatrix}
\begin{bmatrix}
\mathbf{c}_{A} \\
\mathbf{c}_{B}
\end{bmatrix} \\
&= \frac{i}{4}
\begin{bmatrix}
\mathbf{c}_{A}^T & \mathbf{c}_{B}^T
\end{bmatrix}
\begin{bmatrix}
U & 0 \\
0 & V
\end{bmatrix}
\begin{bmatrix}
0 & S \\
-S & 0
\end{bmatrix}
\begin{bmatrix}
U & 0 \\
0 & V
\end{bmatrix}^T
\begin{bmatrix}
\mathbf{c}_{A} \\
\mathbf{c}_{B}
\end{bmatrix}
\equiv \frac{i}{4}
\begin{bmatrix}
\mathbf{c}_{A}^T & \mathbf{c}_{B}^T
\end{bmatrix}
Q
\begin{bmatrix}
0 & S \\
-S & 0
\end{bmatrix}
Q^T
\begin{bmatrix}
\mathbf{c}_{A} \\
\mathbf{c}_{B}
\end{bmatrix} \\
&= \frac{i}{4}
\begin{bmatrix}
(\mathbf{b'})^T & (\mathbf{b''})^T
\end{bmatrix}
\begin{bmatrix}
0 & S \\
- S & 0
\end{bmatrix}
\begin{bmatrix}
\mathbf{b'} \\
\mathbf{b''}
\end{bmatrix}
= \frac{i}{2} \sum_{m=1}^N \varepsilon_m b'_m b''_m = \sum_{m=1}^N \varepsilon_{m} \left(a_m^\dagger a_m - \frac{1}{2} \right),
\end{align}
where $\mathbf{c}_s = \left[ c_{1,s}, c_{2,s}, ..., c_{N,s} \right]^T$ for $s=A,B$. $b'_m$, $b''_m$ are the normal modes, and $a_m^\dagger = \frac{1}{2} \left( b'_m - i b''_m \right)$, $a_m = \frac{1}{2} \left( b'_m + i b''_m \right)$ are the creation and annihilation operators of the complex matter fermion.
The special orthogonal transformation $Q =
\begin{bmatrix}
U & 0 \\
0 & V
\end{bmatrix}$
is derived from the singular value decomposition of $M = U S V^T$ \cite{Vojta_Kitaev_physical_state}.
The matter fermion excitation energies $\varepsilon_m \geq 0$ are the singular values $S = \text{diag}(\varepsilon_1, ... , \varepsilon_N)$, and the $2^N$-dimensional matter fermion Hilbert space $\mathcal{H}_M$ is spanned by $\prod_{m=1}^N \left( a_m^\dagger \right)^{n_m} |M_0\rangle$, $n_m = 0, 1$ with the matter vacuum $|M_0\rangle$ satisfying $a_m |M_0\rangle = 0$ for all $m$.
The matter vacuum and the creation and annihilation operators are implicitly dependent on the background gauge field configuration $\left \{ u^\alpha_{\langle jk \rangle} \right \}$ or equivalently $\left \{ n_{\langle jk \rangle} \right \}$.

It is important to note that eigenstates of the gauge-fixed Hamiltonian $\widetilde{H}_u$ are \emph{not} physical eigenstates of the original Hamiltonian $\hat{H}$.
After we represent the spin operators with Majorana fermions, we enlarged the $2^{2N}$-dimensional Hilbert space $\mathcal{H}_\text{phys}$ for the spins $\hat{\sigma}_j^\alpha$ to the $4^{2N}$-dimensional Hilbert space $\mathcal{H}_F \otimes \mathcal{H}_M$ for the bond fermions $\chi^\alpha_{\langle jk \rangle}$ in $\mathcal{H}_F$ and the matter fermions $a_m$ in $\mathcal{H}_M$.
So not every wavefunction in the enlarged Hilbert space can be physically relevant.
The gauge theory $\widetilde{H}$ is a faithful representation of $\hat{H}$ only if we restrict the theory to the gauge invariant subspace, i.e., a physical wavefunction $|\Psi_\text{phys}\rangle$ needs to be \emph{gauge invariant} such that $\hat{D}_j |\Psi_\text{phys}\rangle = + |\Psi_\text{phys}\rangle$ under a $\mathbb{Z}_2$ gauge transformation $\hat{D}_{j} = b_{j}^x b_{j}^y b_{j}^z c_{j}$ for every site $j$ \cite{Kitaev_anyon}.
Since the gauge transformation $\hat{D}_{j}$ flips sign of the gauge fields $\hat{u}_{jk}^{x,y,z} \to -\hat{u}_{jk}^{x,y,z}$ and the itinerant Majorana fermion $c_{j} \to - c_{j}$ at the site $j$, the eigenstates of the gauge-fixed Hamiltonian are not gauge-invariant.
To get the physical wavefunction $ |\Psi_\text{phys}\rangle$, we must project the gauge-fixed eigenstate $|\psi_u\rangle$ to the physical Hilbert space using the projection operator $\hat{P} = \hat{S} \hat{P}_0$:
\begin{align}
|\Psi_\text{phys}\rangle &= \hat{P} |\psi_u\rangle = \left( \frac{1}{\sqrt{2^{2N-1}}}\sum_g \prod_{j\in g} \hat{D}_j \right) \left(\frac{1+\hat{D}}{2} \right) |\psi_u \rangle
\equiv  \hat{S} \hat{P}_0 |\psi_{u} \rangle,
\end{align}
where $g$ is a collection of all possible proper subset of the lattice and $\hat{D} = \prod_j \hat{D}_j$. Here $\hat{S}$ symmetrically sums over all $2^{2N-1}$ number of physically equivalent eigenstates $|\psi_{u'}\rangle$ of differently gauge-fixed Hamiltonians $\widetilde{H}_{u'}$, and $\hat{P}_0$ projects out all unphysical states $ |\Psi_\mathrm{unphys}\rangle$ with $\hat{D}_j |\Psi_\mathrm{unphys}\rangle = - |\Psi_\mathrm{unphys}\rangle$ for some $j$. Among $4^{2N} = 2^{2N} \times 2^{2N-1} \times 2$ degrees of freedom in $\mathcal{H}$, $2^{2N-1}$ are gauge copies of the same physical state, and only a half of the degrees of freedom with $\hat{D_j} = +1$ are physical.

It has been shown that we can express $\hat{P}_0 = \frac{1}{2}\left( 1 + \hat{D} \right)$ in terms of matter fermion parity, $\hat{\pi} = (-1)^{\sum_m a^\dagger_m a_m}$,
\begin{align}
\hat{D} = (-1)^\theta \left( \prod_{\langle jk \rangle} u^\alpha_{jk} \right) \left( \prod_\mu i c_{\mu,A} c_{\mu,B} \right)
= (-1)^\theta \left(\mathrm{det} \,Q \prod_{\langle jk \rangle} u^\alpha_{jk} \right) \hat{\pi},
\end{align}
where $(-1)^\theta$ is determined by the boundary condition, and $Q$ is the special orthogonal transformation to diagonalize $\widetilde{H}_u$ \cite{Pedrocchi_Kitaev_physical_state}.
Since the gauge transformation $\hat{D}_j$ flips the sign of $\hat{u}_{jk}^\alpha$ and $c_j$,  the parity of $b$ fermions (which is equal to the parity of the bond fermions $\chi^\alpha$, $\hat{\pi}_b = \hat{\pi}_\chi = \prod_{\langle jk \rangle} u^\alpha_{jk}$) and the parity of $c$ fermions ($\hat{\pi}_c = \prod_\mu i c_{\mu,A} c_{\mu,B}$) are not gauge invariant.
However, the parity of the complex matter fermion $\hat{\pi}$ must be gauge invariant because the occupation number of $a_m$ determines the density of states which directly influences physical observables. Since $\hat{D}$ is also gauge invariant, we can deduce which matter fermion parity $\hat{\pi}$ is physical for a given flux sector by calculating $\mathrm{det} \,Q \,\hat{\pi}_\chi$.
With the standard gauge choice $u^\alpha_{jk} = +1$ for the zero flux sector, two-flux states and four-flux states can be created by flipping appropriate one and two $z$-bonds, respectively.
Hence, the zero-flux and four-flux states have the same $\chi$ fermion parity $\hat{\pi}_\chi$, but the two-flux states have the opposite $\hat{\pi}_\chi$.
We numerically calculated $\mathrm{det} \,Q$ for various system sizes and found that the zero-, two-, and four-flux states have the same value of $\mathrm{det} \,Q$.
Therefore the zero-flux and four-flux states should have the same matter fermion parity $\hat{\pi}$, but the two-flux states have the opposite matter fermion parity \cite{Vojta_Kitaev_physical_state}.

\section{Matrix elements for spin operators}

To calculate the four-point functions, the matrix elements of $\hat{\sigma}^z_j$ between the energy eigenstates are necessary. Let $|R_\mu \rangle$ be the two-flux states with one matter fermion $\overline{a}_r$ having energy $\varepsilon_r$, i.e.,
$|R_\mu \rangle = \hat{P} |F_2\rangle(\overline{a}_r^\dagger |\overline{M}_0\rangle) = \hat{P}\chi_\mu^z\overline{a}_r^\dagger |F_0\rangle |\overline{M}_0\rangle$,
where $\chi_\mu^z$ is the bond fermion at the $z$-bond of $\mu$th unit cell, and $|\overline{M}_0\rangle$ is the matter vacuum of $\widetilde{H}_{\overline{u}}$ with two-flux gauge configuration $\overline{u}$. The matrix element of the spin operator $\hat{Z}_\mu = \hat{\sigma}_{\mu,A}^z + \hat{\sigma}_{\mu,B}^z$ between the two-flux state $|R_\mu\rangle$ and the ground state $|G\rangle = \hat{P}|F_0\rangle |M_0\rangle$ is calculated in Ref.~\cite{Vojta_Kitaev_physical_state}:
\begin{align}
\langle R_\mu | \hat{\sigma}_{\mu,A}^z + \hat{\sigma}_{\mu,B}^z | G \rangle
&\equiv \langle R_\mu | \hat{Z}_{\mu} | G \rangle
= \langle \overline{M}_0 | \langle F_0 | \overline{a}_r \chi_\mu^{z\dagger} \hat{P}  \hat{Z}_{\mu} \hat{P} |F_0\rangle |M_0\rangle \\
&= \langle \overline{M}_0 | \langle F_0 | \overline{a}_r \chi_\mu^{z\dagger} \left[ i (\chi_\mu^z + \chi_\mu^{z\dagger})c_{\mu,A} +(\chi_\mu^z - \chi_\mu^{z\dagger}) c_{\mu,B} \right]
(1 + \sum_j D_j + \cdots ) |F_0\rangle |M_0\rangle \\
&= \langle \overline{M}_0 | \overline{a}_r ( i c_{\mu,A} + c_{\mu,B}) |M_0\rangle.
\end{align}
The matter fermion annihilation operator $\overline{a}_r$ and its vacuum $|\overline{M}_0\rangle$ in the two-flux sector can be written in terms of the creation and annihilation operator $a_r^\dagger$, $a_r$ of the matter fermions and the matter vacuum $|M_0\rangle$ in the zero-flux sector.
With singular value decompositions of the two-flux Hamiltonian $\widetilde{H}_{\overline{u}} = \overline{U}_2 \overline{S}_2 \overline{V}_2^T$ and the zero-flux Hamiltonian $\widetilde{H}_u = U_0 S_0 V_0^T$,
\begin{align}
|\overline{M}_0\rangle &= \sqrt{|\mathrm{det}\,\overline{X}|} e^{-\frac{1}{2} a_m^\dagger \overline{F}_{mn} a_{n}^\dagger} |M_0\rangle, \label{eq:M0connect}\\
\overline{a}_r &= \overline{X}_{rr'}a_{r'} + \overline{Y}_{rr'}a^\dagger_{r'}, \label{eq:a_connect}
\end{align}
where Einstein's summation convention is assumed for the repeated Latin indices labelling the energy eigenstates, and $\overline{X} = \frac{1}{2}( \overline{U}_2^T U_0 + \overline{V}_2^T V_0)$, $\overline{Y} = \frac{1}{2}( \overline{U}_2^T U_0 - \overline{V}_2^T V_0)$, $\overline{F} = (\overline{X})^{-1}\overline{Y}$.
The $c$ Majorana fermion can be also expressed in terms of the zero-flux matter fermion,
\begin{align}
c_{\mu,A} = (U_0)_{\mu,s}(a_s + a_s^\dagger),
~c_{\mu,B} = i(V_0)_{\mu,s}(a_s ^\dagger - a_s).
\end{align}
Therefore
\begin{align}
\langle R_\mu | \hat{Z}_{\mu} | G \rangle &= \sqrt{|\mathrm{det}\,\overline{X}|}\langle M_0 | (\overline{X}_{rr'}a_{r'} + \overline{Y}_{rr'}a^\dagger_{r'} ) ( i (U_0 + V_0)_{\mu,s} a_s^\dagger + i(U_0 - V_0)_{\mu,s} a_s ) |M_0\rangle \\
&= i \sqrt{|\mathrm{det}\,\overline{X}|} \left[ (U_0 + V_0)(\overline{X})^{-1}\right]_{\mu,r}
\label{eq:F2M1_Z0_F0M0}
\end{align}

Similarly, we can define the four-flux matter vacuum state $|Q_{\mu\nu}\rangle = \hat{P} \chi_\mu^z \chi_\nu^z |F_0\rangle |M_0'\rangle$.
By following the similar calculations in Ref.~\cite{Vojta_Kitaev_physical_state, Blaizot_quantum_theory},
\begin{align}
\langle R_\nu | \hat{\sigma}_{\mu,A}^z + \hat{\sigma}_{\mu,B}^z | Q_{\mu\nu}\rangle
&\equiv \langle R_\nu | \hat{Z}_{\mu} | Q_{\mu\nu}\rangle
= \langle \overline{M}_0 | \langle F_0 | \overline{a}_r \chi_\nu^{z\dagger} \hat{P}  \hat{Z}_{\mu} \hat{P}  \chi_\mu^z \chi_\nu^z |F_0\rangle |M_0'\rangle \\
&= \langle \overline{M}_0 | \langle F_0 | \overline{a}_r \chi_\nu^{z\dagger} \left[ i (\chi_\mu^z + \chi_\mu^{z\dagger})c_{\mu,A} +(\chi_\mu^z - \chi_\mu^{z\dagger}) c_{\mu,B} \right]
(1 + \sum_j D_j + \cdots ) \chi_\mu^z \chi_\nu^z |F_0\rangle |M_0'\rangle \\
&= \langle \overline{M}_0 | \overline{a}_r ( i c_{\mu,A} - c_{\mu,B}) |M_0'\rangle
= \cdots = i \sqrt{|\mathrm{det}\, X'|} \left[ (U'_4 - V'_4)(X')^{-1}\right]_{\mu,r},
\label{eq:F2M1_Zl_F4M0}
\end{align}
where $U_4'$ and $V_4'$ are the left and right singular matrices of the four-flux Hamiltonian $\widetilde{H}_{u'} = U_4' S_4' {V_4'}^T$, and $X' = \frac{1}{2} (\overline{U}_2^T U_4' + \overline{V}_2^T V_4')$. 
Note that $|Q_{\mu\nu}\rangle = - |Q_{\nu\mu}\rangle$ implies
\begin{align}
\langle R_\mu | \hat{\sigma}_{\nu,A}^z + \hat{\sigma}_{\nu,B}^z | Q_{\mu\nu}\rangle
&\equiv \langle R_\mu | \hat{Z}_{\nu} | Q_{\mu\nu}\rangle
= - \langle R_\mu | \hat{Z}_{\nu} | Q_{\nu\mu}\rangle
= - i \sqrt{|\mathrm{det}\, X'|} \left[ (U_4' - V_4')(X')^{-1}\right]_{\nu,r}.
\label{eq:F2M1_Z0_F4M0}
\end{align}

\section{Real-time four-point correlation functions} 
\begin{figure} [h]
\centering
\includegraphics[width=\textwidth]{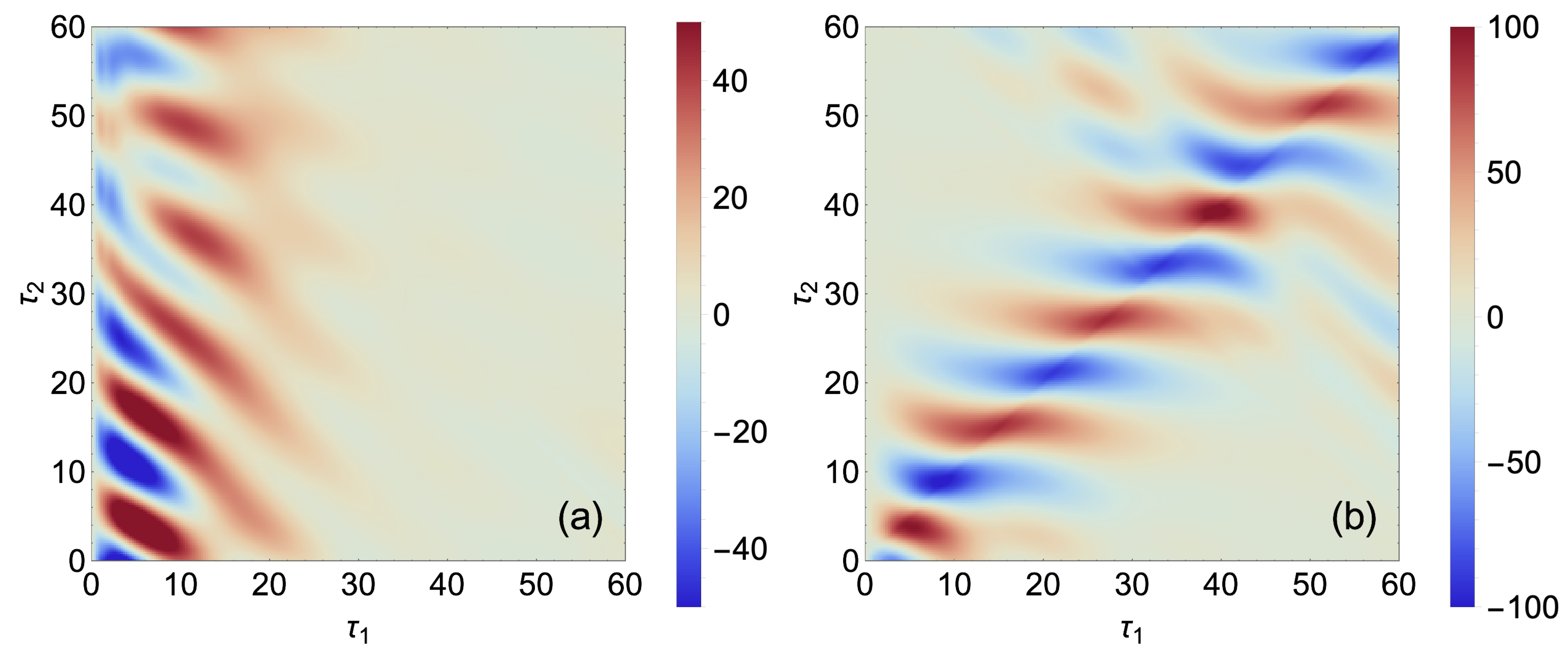}
\caption{Real-time third-order susceptibilities (a) $\chi^{(3),z}_{zzz}(\tau_2,\tau_1,0)$ (b) $\chi^{(3),z}_{zzz}(\tau_2,0,\tau_1)$}
\label{fig:chi_t}
\end{figure}

We calculated the third-order nonlinear susceptibilities from the four-point correlation functions, which are mostly out of time order (Fig.~\ref{fig:chi_t}). 
\begin{figure} [h]
\centering
\includegraphics[width=\textwidth]{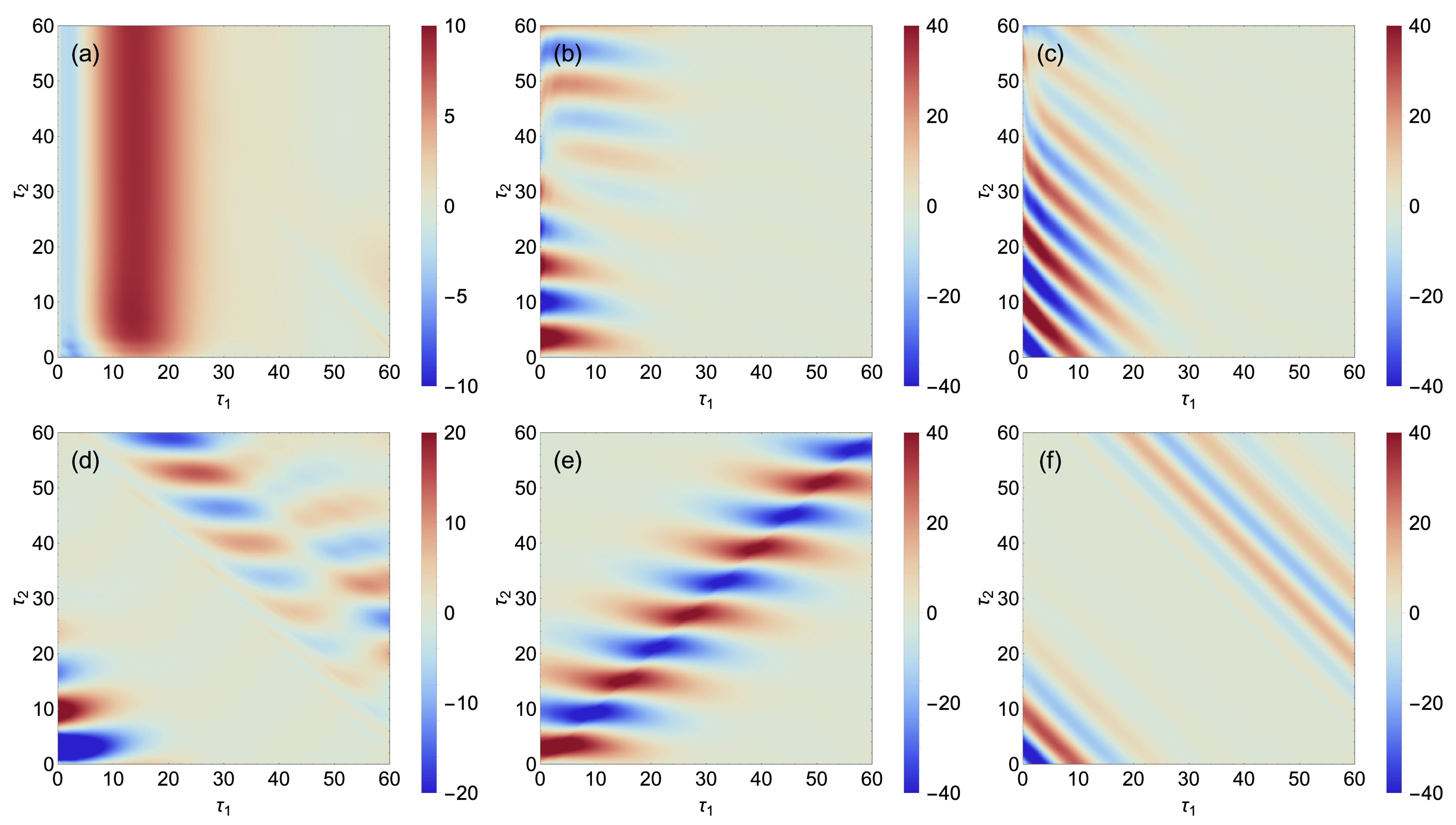}
\caption{Real-time dynamical correlation functions for the isotropic Kitaev model on a $125 \times 125$ honeycomb lattice. \newline
(a) $R^{(1,2),z}_{zzz}(\tau_2,\tau_1,0)$ (b) $R^{(3),z}_{zzz}(\tau_2,\tau_1,0)$ (c) $R^{(4),z}_{zzz}(\tau_2,\tau_1,0)$ (d) $R^{(1),z}_{zzz}(\tau_2,0,\tau_1)$ (e) $R^{(2,3),z}_{zzz}(\tau_2,0,\tau_1)$ (f) $R^{(4),z}_{zzz}(\tau_2,0,\tau_1)$}
\label{fig:Rt}
\end{figure}

Using the resolution of identity $\sum_{P} |P\rangle \langle P| = 1$ with the energy eigenstate $|P\rangle$, we decompose each four-point function into a sum of products of four matrix elements of the spin operator. For example,
\begin{align}
R^{(1),z}_{zzz}(\tau_2,\tau_1,0)  &= \langle G|\hat{M}^{z} (0) \hat{M}^{z}(\tau_1)\hat{M}^{z}(\tau_2+\tau_1)\hat{M}^{z}(0) |G \rangle \nonumber \\
&= \sum_{PQR} \langle G | \hat{M}^{z} | P \rangle \langle P| e^{i\hat{H}\tau_1} \hat{M}^z e^{-i\hat{H}\tau_1} |Q\rangle
\langle Q|e^{i\hat{H}(\tau_2+\tau_1)} \hat{M}^z e^{-i\hat{H}(\tau_2+\tau_1)} |R\rangle \langle R| \hat{M}^z |G\rangle \\
&= \sum_{PQR} e^{i (E_P-E_Q)\tau_1 +i(E_Q-E_R)(\tau_2+\tau_1)} \langle G|\hat{M}^z|P\rangle\langle P|\hat{M}^z|Q\rangle
\langle Q|\hat{M}^z|R\rangle\langle R|\hat{M}^z|G\rangle \\
&=\sum_{\mu\nu\lambda\rho} \sum_{Q} e^{iE_Q \tau_2}
\sum_P e^{iE_P \tau_1} \langle G|\hat{Z}_\mu |P\rangle\langle P|\hat{Z}_\nu |Q\rangle
\sum_R e^{-iE_R(\tau_2+\tau_1)} \langle Q|\hat{Z}_\lambda |R\rangle\langle R|\hat{Z}_\rho |G\rangle \label{eq:genR1z}
\end{align}
Since each operator $\hat{Z}_\mu = \hat{\sigma}^z_{\mu,A} + \hat{\sigma}^z_{\mu,B}$ at a $z$-bond of the $\mu$th unit cell excites two fluxes adjacent to the $z$-bond, we can infer the flux sectors of $|P\rangle$, $|Q\rangle$, and $|R\rangle$ so that
\begin{align}
R^{(1),z}_{zzz}(\tau_2,\tau_1,0) &=N \sum_\mu e^{i(E_0 - E_2)\tau_2}\sum_{Q} e^{i\varepsilon_Q \tau_2}
\sum_P e^{i\varepsilon_P \tau_1} \langle G|\hat{Z}_\mu |P\rangle\langle P|\hat{Z}_\mu |Q\rangle
\sum_R e^{-i\varepsilon_R(\tau_2+\tau_1)} \langle Q|\hat{Z}_0 |R\rangle \langle R|\hat{Z}_0 |G\rangle \label{eq:R1a0} \\
&+N\sum_{\mu\neq0} e^{i(E_4(\mu) - E_2)\tau_2} \sum_{Q} e^{i\varepsilon_Q \tau_2}
\sum_P e^{i\varepsilon_P \tau_1} \langle G|\hat{Z}_0 |P\rangle\langle P|\hat{Z}_\mu |Q\rangle
\sum_R e^{-i\varepsilon_R(\tau_2+\tau_1)}
\langle Q|\hat{Z}_\mu |R\rangle\langle R|\hat{Z}_0 |G\rangle \label{eq:R1a4i} \\
&+N\sum_{\mu\neq0} e^{i(E_4(\mu) - E_2)\tau_2} \sum_{Q} e^{i\varepsilon_Q \tau_2}
\sum_{\widetilde{P}} e^{i\varepsilon_{\widetilde{P}} \tau_1} \langle G|\hat{Z}_\mu |\widetilde{P} \rangle\langle \widetilde{P}|\hat{Z}_0 |Q\rangle
\sum_R e^{-i\varepsilon_R(\tau_2+\tau_1)} \langle Q|\hat{Z}_\mu |R\rangle\langle R|\hat{Z}_0 |G\rangle, \label{eq:R1a4ii}
\end{align}
where $E_0$ is the energy of the ground state $|G\rangle$, and $\varepsilon_{P}$, $\varepsilon_{Q}$, $\varepsilon_{R}$ are the total matter fermion excitation energies of $|P\rangle$, $|Q\rangle$, and $|R\rangle$, respectively.
$E_0$, $E_2$, $E_4(\mu)$ are the energies of the matter fermion vacuum states with the zero-flux, two-flux, and four-flux background.
Note that $E_4(\mu)$ depends on the relative distance between two pairs of fluxes; two fluxes share the $z$-bond at the 0th unit cell and the other two flux share the $z$-bond at the $\mu$th unit cell.
Using translational symmetry of the zero flux states, we simplified the summation $\sum_\rho \to N \sum_\rho \delta_{\rho,0}$.
Similarly, we can decompose and reorganize the other correlation functions $R^{(l=2,3,4)}(\tau_2,\tau_1,0)$ and $R^{(l=1,2,3,4)}(\tau_2,0,\tau_1)$.

After we approximate the correlation function by truncating the summation $\sum_{PQR}$ up to one-matter-fermion states, we calculated the four-point functions from $\tau_{1,2} = 0$ to $\tau_{1,2} = 60$ with $d\tau = 0.25$ (Fig.~\ref{fig:Rt}).
To calculate the matrix elements, Eqs.~(\ref{eq:F2M1_Z0_F0M0}), (\ref{eq:F2M1_Zl_F4M0}), and (\ref{eq:F2M1_Z0_F4M0}), singular value decompositions (SVDs) of $N\times N$ matrices, of which calculation complexity is $\mathcal{O}(N^3) =\mathcal{O}(L^6)$, are necessary for $N = L \times L$ unit-cell cluster of the honeycomb lattice.
For Eqs.~(\ref{eq:R1a4i}) and (\ref{eq:R1a4ii}), $N - 1$ number of all different four-flux configurations $|Q\rangle$ must be summed over because of $\sum_{\mu \neq 0}$.
For a given $\mu$, two different SVDs are necessary for the two-flux state $|\widetilde{P}\rangle$ and the four-flux state $|Q\rangle$. 
Therefore the computational cost quickly grows as $\mathcal{O}(L^8)$ even after the summation is truncated.
Ripple-like anomalous long time behavior is finite size effect \cite{Wan_IsingTHz}, so we truncate the real-time correlation functions and use the fast Fourier transformations (FFT) to obtain the two-dimensional Fourier spectra for the third-order susceptibilities.
The diagonal signals in the first and third quadrants originating from $R^{(1)}(\tau_2,0,\tau_1)$ and $R^{(4)}(\tau_2,0,\tau_1)$ [Fig. 2 (d) and (f) of the Letter] are sensitive to how we truncate the two-dimensional real-time susceptibilities before the Fourier transformations. However, the sharp diagonal signals in the second and fourth quadrants due to $R^{(3)}(\tau_2,0,\tau_1)$ [Fig. 2 (e) of the Letter]  are robust and hardly dependent to the way we crop the real-time data.


\end{document}